\input{epsf}

\documentclass[preprint,showpacs,preprintnumbers,amsmath,amssymb]{revtex4}

\usepackage{graphicx}

\begin{document}
\draft
Published as: Ehrlich, R., Astropart. Phys., {\bf 35}, 625-628 (2012). 
\\

\title{Evidence for two neutrino mass eigenstates from SN 1987A and the possibility of superluminal neutrinos
}

\author{Robert Ehrlich, rehrlich@gmu.edu}
\address{Physics, Astronomy and Computational Sciences\\
George Mason University\\ Fairfax, VA 22030\\
703-993-1268\\703-993-1269(FAX)}

\begin{abstract}
This paper reports a new phenomenological analysis of the neutrino burst detected from SN 1987 A, and it reveals the presence of two mass eigenstates.  The heavier mass eigenstate has  $m_H=21.4 \pm 1.2 eV/c^2$, while the lighter one has  $m_L=4.0 \pm0.5 eV/c^2 $.   It is not the first paper to make such a claim, but it expands on a 1988 conditional analysis by Cowsik, and it attempts to make the evidence more robust through an improved statistical analysis, and through providing reasons why alternative explanations are unlikely.  It also shows how the result can be made consistent with existing smaller electron neutrino mass limits from tritium beta decay and cosmology through the existence of a third tachyonic (superluminal) mass eigenstate.\end{abstract}


\maketitle

\section{Introduction}
The neutrinos from SN 1987A have been the subject of hundreds of papers, both theoretical and phenomenological.\cite{Pagliaroli}  Some of these papers analyze the data to infer an upper limit on the neutrino mass, which ranges typically from 12 to 16 $eV/c^2$.\cite{Bahcall, Arnett}.   It is expected however that neutrinos from a supernova should include all three active neutrino mass eigenstates, therefore such a limit must represent some sort of average over whichever eigenstates are represented in the neutrino events recorded.   There have in fact been at least three analyses of the SN 1987A neutrino data that have given evidence for a bifurcation into two groupings.\cite{Cowsik, Loredo, Roos}.  The analysis by Loredo was based on Baysian statistics,\cite{Loredo} and it gave no physical interpretation to the two classes of events.  The Roos paper based the bifurcation on neutrino mass, but he attributed the different mass values to the events recorded in the two detectors, rather than to two mass eigenstates.  Finally, the Cowsik paper did suggest a bifurcation associated with two mass eigenstates whose masses he found based on the relation between neutrino arrival times and their measured energies.  Cowsik reported these eigenstates as $24 \pm 7 eV/c^2$ and  $4 \pm 1 eV/c^2$.  We here extend his analysis to include the possibility of observed regularity arising in part from chance, or variations in their emission times at the source, since his conclusion was conditional on there being near instantaneous emissions.  Additionally, we show how the result can be made consistent with existing electron neutrino mass limits, and other experiments.  As with most SN 1987A analyses, we include neutrino events recorded in the Kamioka, and IMB detectors, but exclude the 5 events seen in the Mont Blanc detector since they occured 5 hours before those detected by the other three detectors.  Additionally, given its size and sensitivity that detector should have only seen 1 neutrino from the supernova not 5.   We also exclude 5 events seen in the Baksan detector because there are substantial reasons to believe that perhaps 2 or 3 of them may have been due to background.  

As noted in the paper on the Baksan neutrinos from SN 1987A.\cite{Baksan}

\begin{itemize}
\item The 5 event burst they saw over a 10 s period was only slightly above their background, since there have been past instances when the Baksan detector recorded 3 "random" events in 10 seconds
\item The timing of the Baksan burst occurred 30 s after those seen in the IMB and Kamioka detectors
\item The 5 Baksan events were spread evenly over the entire 10 s period, not bunched strongly near the start of the 1987A burst as with the other detectors.  Thus, there is considerable uncertainty over the true $t=0$ time for these events relative to the other two detectors.
\end{itemize}

The basic idea of the analysis is straightforward.  A neutrino of mass m and energy E that is emitted at a time $t_{em}$ and observed in a detector at time t will take a time to reach Earth given by

\begin{equation}
\Delta t= t -t_{em}\approx t_0(1+m^2/2E^2)
\end{equation}

where the approximation requires that $|m^2|$ be small compared to $E^2$ and where $t_0$ is the light travel time from SN 1987A or 168,000 years.  Eq. 1 allows us to calculate a mass $m_k$ for the kth neutrino event observed at a time $t_k$ whose energy is $E_k$, assuming the neutrino was emitted at time $t_{em,k}$, and travelled a distance $ct_{0,k}$ to reach Earth.  Note that $t_{0,k}$ can vary from one event to another depending on the collapsing core radius at the time that neutrino was emitted.  On solving for the mass attributed to a given event, we have:

\begin{equation}
m^2_k =\frac{2E^2(t_k-t_{em,k}-t_{0,k})}{t_{0,k}}
\end{equation}

but clearly eq. 2 can only be used to find $m_k$ when $t_{em,k}$ and $t_{0,k}$ are both nearly event-independent, which requires near-simultaneous emissions.  

\section{Evidence for near-simultaneous emissions}

The evidence for near simultaneous emissions comes from supernova core collapse models and their calculated neutrino fluxes as a function of time, specifically:  

\begin{enumerate} 
\item \emph{brief "burst" of $\nu_e$}  
The literature on electron neutrinos and antineutrinos emitted during a core collapse consistently shows that both fluxes rise and fall over a very short time interval.  Typically, this initial burst is found to rise and fall by almost an order of magnitude over a time interval of 0.3 seconds,\cite{Bruenn, Pagliaroli} while some models show this "burst" of electron neutrinos and antineutrinos being as short as about 0.02 seconds.\cite{Myra}  

\item \emph{greater $\nu_e$ than $\nu_\mu$ flux during burst}
These core collapse simulations find that the drop in emitted muon and tau neutrinos following its peak is much more gradual and extends over $~10-15 sec$, but the initial  $\nu_e$  flux during its short "burst" is found to be about an order of magnitude greater than the $\nu_\mu$  flux for $t>1s$,\cite{Myra} and in any case, the SN 1987A neutrinos observed are unlikely to be these flavors -- certainly not for the IMB detector whose threshold was below muon neutrino detection. 

\item \emph{possibility of brief burst for $\nu_\mu$}
Although some core collapse simulations find that the drop in emitted muon and tau neutrino fluxes following its peak is much more gradual than $\nu_e$, other authors suggest that a similar rapid rise and fall applies to these two flavors as well.\cite{Bratton} -- see Keil, et.al for a summary of predictions from various models.\cite{Keil} 

\item \emph{correlation between neutrino energy and emission time}
If the later arriving neutrinos are simply late because they were emitted at later times, one would expect them to have higher energies than the earlier emitted neutrinos according to the models, but this is not what the data show.

\end{enumerate}

Given the above, it is not unreasonable to assume that most (or all) of the SN1987A neutrinos detected were emitted within a $\Delta t \approx 0.3 s$ time interval.  Moreover, given the rapidity of the core collapse (whose radius drops by an order of magnitude or more depending on mass) within a time of 0.2 sec,\cite{Bruenn-DeNisco} we may assume that the neutrino travel distance changes very little for those emitted at the start and end of the $0.3 s$ burst.  Hence, taking $t_{em}=-t_0$, we may plausibly assume that within roughly a $\pm 0.2$ sec uncertainty $t_{em,k}+t_{0,k}\approx t_{em}+t_0=0\pm 0.2 sec.$

Thus, within this time uncertainty which we assign to the observed $t_k's$, Eq. 2 becomes:

\begin{equation}
m^2_k =\frac{2E^2t_k}{t_0}
\end{equation}

\begin{figure}
\epsfxsize=350pt 
\epsffile{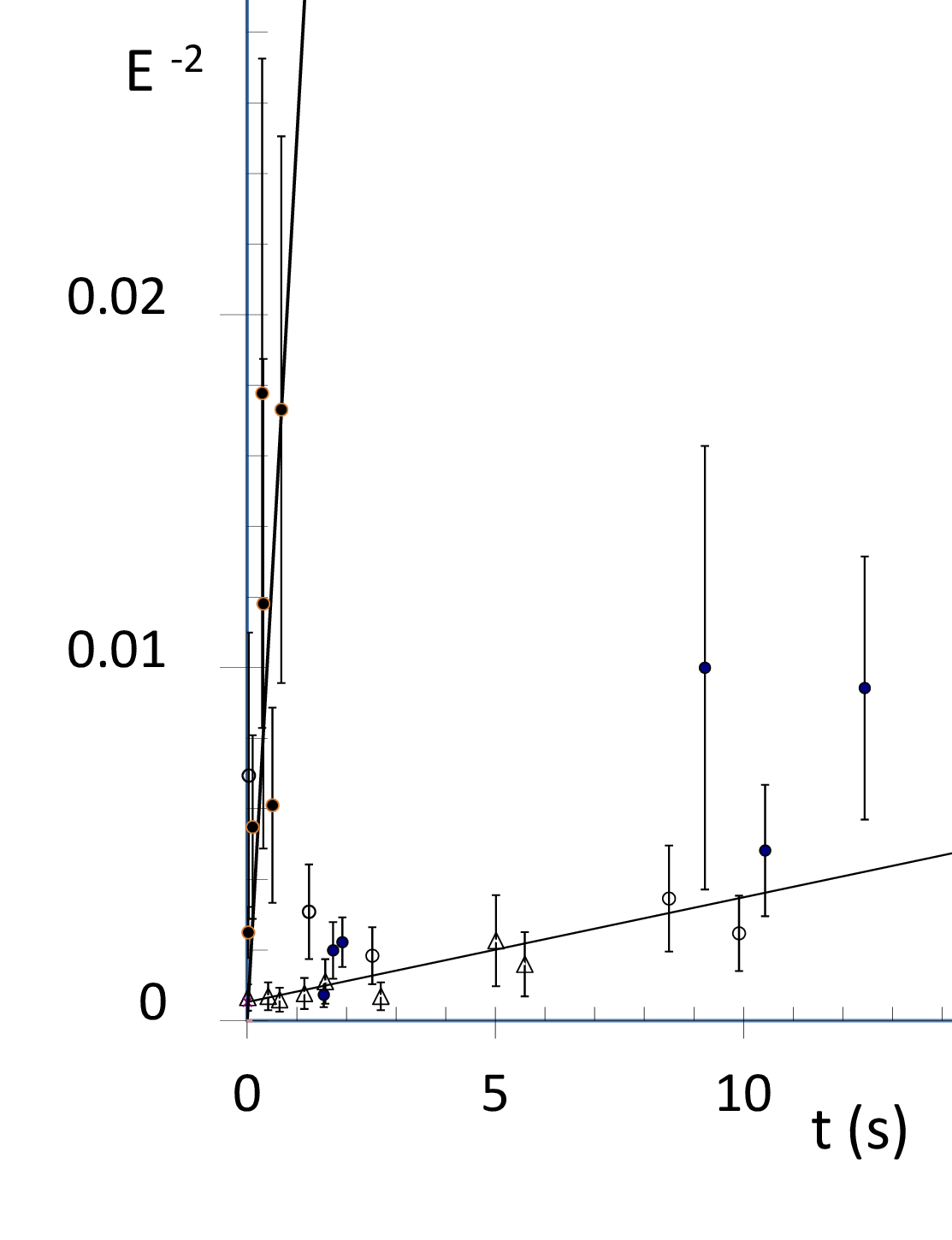} \\
\caption{\small On a plot of $E^{-2}$ versus t, the SN 1987A data all lie close to one of two straight lines that nearly pass through the origin.  Solid circles show Kamioka events, triangles IMB events, and open circles Baksan events, which were not used in the analysis to find masses indicative of two distinct masses.
 \\}
\end{figure}

We note that according to Eq. 3 $t=0$ for any $E$ requires $m=0$.  However, for a pair of unsynchronized detectors, for which $t=0$ was arbitrarily chosen to mark the arrival of their first event, it is unlikely that this time corresponds to the instant a photon emitted with the neutrino burst would have reached the detector -- especially when the number of recorded events in each detector is small.  However, straight line fits of $1/E_k^2$ versus $t_k$ show that the data is consistent with one of two lines both cosnsistent with passing through the origin.  Therefore, apart from a small time offset for each detector $t=0$ is consistent with $t_{em}=-t_0$.  For our estimate of the size of this offset in each detector we have chosen 1/4 the time difference between each detector's first and second events: 0.107 s in Kamioka, and 0.42 s in IMB in using Eq. 3 to find the mass for each event.  However, since these estimated detector offsets are rather uncertain, the $\pm 0.2s$ uncertainty due to a spread in emission times at the source has been increased to a net uncertainty for all the $t_k$ of $\sigma_t= 0.5s$ when computing the uncertainty in the mass associated with each event using:  

\begin{equation}
 \sigma (m^2_k) = \sqrt{\sigma^2 (E_k^2) + \sigma^2 (t_k)}
\end{equation}
and $\sigma_k= \sigma (m_k^2)/2m_k$

\begin{table}[hbt]
\begin{center}
\begin{tabular}{ l c r rrrrr r r r}

\hline
Event && E in MeV &&&&&& t in s && mass\\
\hline
K7             &&  $36.9\pm 8.0$     &&&&&& 1.541     && $28.1\pm 5.6$ \\
K8             &&  $22.4\pm 4.2$     &&&&&& 1.728     && $18.1\pm 3.2$ \\
K9            &&  $21.2\pm 3.2$     &&&&&& 1.915     && $18.0\pm 2.6$ \\
K10           &&  $10.0\pm 2.7$     &&&&&& 9.219     && $18.7\pm 4.5$ \\
K11           &&  $14.4\pm 2.6$     &&&&&& 10.43     && $28.6\pm 4.8$ \\
K12          &&  $10.3\pm 1.9$     &&&&&& 12.44     && $22.3\pm 3.8$ \\
I2            &&  $37.0\pm 9.25$     &&&&&& 0.42     && $14.7\pm 3.6$ \\
I3            &&  $40.0\pm 10.0$     &&&&&& 0.65     && $19.8\pm 4.6$ \\
I4            &&  $35.0\pm 8.75$     &&&&&& 1.15     && $23.1\pm 5.3$ \\
I5            &&  $29.0\pm 7.25$     &&&&&& 1.57     && $22.3\pm 5.1$ \\
I6            &&  $37.0\pm 9.25$     &&&&&& 2.69     && $37.3\pm 8.4$ \\
I7            &&  $20.0\pm 5.0$     &&&&&& 5.01     && $27.5\pm 6.2$ \\
I8            &&  $22.4\pm 4.2$     &&&&&& 1.728     && $18.1\pm 3.2$ \\
\hline
&&&&&&       $\chi^2 = 17.8$ &&&&    average  mass =  $21.4\pm 1.2$ \\
\end{tabular}
\end{center}
\caption{Masses in $eV/c^2$ for 6 Kamioka events and 7 IMB events that are clearly clustered about a weighted average of 21.4 $eV/c^2$.  Each event's mass is calculated from Eq. 3 using their observed time and energy with their associated propagated errors.}
\end{table}

Tables I and II show the results for the 20 events recorded in the two detectors including 12 \cite{Number} in Kamioka, and  8 in IMB.  For clarity, we have separated the 20 events based on whether they correspond to a mass above (Table I) or below (Table II) $10 eV/c^2.$ There is a very clear clustering of events into two groups with 13 of them all clustering about $m_H = 21.4\pm 1.2 eV/c^2$ and the 7 remaining ones clustering about $m_L = 4.0 \pm 0.5 eV/c^2.$  There is no ambiguity in terms of the cluster to which any one event belongs, since in no case was any event less than $2.9\sigma$ away from the center of the "wrong" cluster, and a majority of events were more than $4\sigma$ away.  Moreover, with each cluster the spread in event masses was very consistent with what one would expect given the size of the mass uncertainties.  Thus, the chi squares for all the events fitting one of two masses were quite acceptable: $\chi ^2_L =5.9$ (6 dof. p=0.43) and $\chi ^2_H =17.8$ (12 dof, p=0.12).  In contrast, if the 20 events were assumed to be described by a single mass, the best value would be $7.9\pm 0.5,$ with $\chi ^2 =203.4$ (19 dof, $p < 10^{-30}$).  One might object to this comparison on the grounds that an alternative null hypothesis to clustering about two masses would be simply a flat distribution of masses, which would in fact give an acceptable chi square.  However, given that it is known that at most three active mass eigenstates are present, such a distribution is physically unrealistic, assuming that the supernova core collapse models are correct in predicting a very small spread in emission times.

The ideogram of Fig. 2 offers a visual way to display the degree of clustering of the 20 events about two specific masses.  In this display each event is represented by a normalized Gaussian $G(m_k,\sigma_k),$ with mean and standard deviation $m_k$ and $\sigma_k$.  In addition to showing the masses of each event with error bars, we also display the value of $\Sigma^{20}_{k=1} G(m_k,\sigma_k)$.

\begin{table}[hbt]
\begin{center}
\begin{tabular}{ l c r rrrrr r r r}

\hline
Event && E in MeV &&&&&& t in s && mass\\
\hline
I1             &&  $38.0\pm 9.5$     &&&&&& 0     && $0.0\pm 7.3$ \\
K1             &&  $21.3\pm 2.9$     &&&&&& 0     && $0.0\pm 4.7$ \\
K2            &&  $14.8\pm 3.2$     &&&&&& 0.107 && $3.0\pm 1.7$ \\
K3           &&  $8.9\pm 2.0$     &&&&&& 0.300     && $3.0\pm 1.0$ \\
K4           &&  $10.6\pm 2.7$     &&&&&& 0.324     && $3.7\pm 1.2$ \\
K5          &&  $14.4\pm 2.9$     &&&&&& 0.507     && $6.3\pm 1.4$ \\
K6            &&  $7.6\pm 1.7$     &&&&&& 0.686     && $3.9\pm 0.9$ \\
\hline
 &&&&&&       $\chi^2 = 5.9$ &&&&    average  mass =  $4.0\pm 0.5$ \\

\end{tabular}
\end{center}
\caption{Masses in $eV/c^2$ for 6 Kamioka events and 1 IMB event that are clustered about a weighted average of 4.0 $eV/c^2$.  Each event's mass is calculated from Eq. 3 using their observed time and energy with their associated propagated errors.}
\end{table}

\section{Discussion of results}
In terms of the robustness of these results, it should be noted that:
\begin{enumerate}

\item \emph{Every} Kamioka or IMB event in the data is included in this analysis, using the published values for t, E and $\Delta E$,\cite{Kamioka, IMB}.  Thus, Baksan aside, it would appear there is no noise or background events in the SN 1987A data.  Moreover unlike earlier interpretations of the data where only events with $t<2-3$ seconds are assumed to be part of the main supernova burst here it is assumed that they all are.

\item \emph{Every} event is cleanly associated with \emph{only} one of the two mass eigenstates given its error bars, and as noted previously the chi squares for the two fits are quite acceptable.  In contrast fitting all 20 events using a single Gaussian distribution gives an infinitessimally small probability.

\item Every event is consistent with $m^2$ values that are positive.   In any case, no assumption was made at the outset in this analysis about whether there were or were not events having "tachyonic" or superluminal masses, which would have to lie on negatively sloped lines in Fig.1.

\item As a general rule it is not very unlikely to find some kind of interesting pattern with only 20 events just based on chance.  However, finding a highly specific pattern (every single event strongly clustering about one of two mass values is quite another matter, especially given the support from supernova models showing that the detection times are very likely indicators of the neutrino travel times within a $\pm 0.5s$ uncertainty in $t_k$.  One could choose to disregard the present models of supernova core collapse, and insist on a significant spread in neutrino emission times.  But in that case, one would need to assume the existence of some kind of weird correlation between the varying emission times and the emitted neutrino energies that perfectly mimics the presence of two discrete mass eigenstates at Earth's particular distance from SN 1987A.  Moreover, this correlation would fail to produce that result if the distance, $L_0$ to SN 1987A happened to be significantly different, because the neutrino travel times for a given energy are proportional to $L_0$, making this alternative explanation extremely improbable.
\end{enumerate}

\begin{figure}
\epsfxsize=300pt 
\epsffile{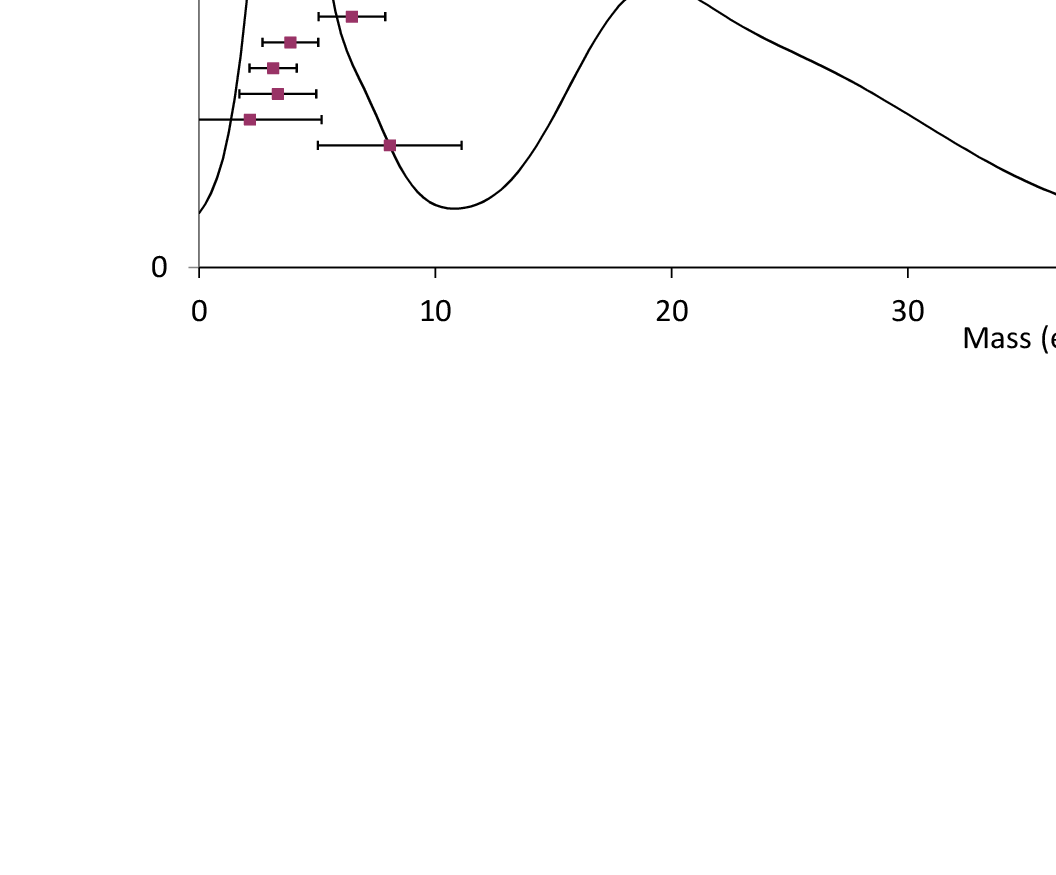} \\
\caption{\small Ideogram of the 20 SN 1987A events computed from the sum of 20 normalized Gaussians, i.e., $\Sigma^{20}_{k=1} G(m_k,\sigma_k)$ which displays the clustering of the events around two distinct masses.
 \\}
\end{figure}

\section{Implausible masses, a missing third eigenstate and OPERA}
The values reported here for the two observed mass eigenstate might seem implausibly high.  Given that the electron neutrino presumably consists of a mixture of the light mass eigenstate plus one or both of the other two, the $\nu_e$ mass seemingly would need to exceed that of $m_L=4.0\pm 0.5 eV/c^2,$ which is inconsistent with the upper limit on the mass of the electron neutrino from tritium decay: $m< 2 eV/c^2$\cite{PDG}.  However, there is an interesting possible explanation of this anomaly that depends on the OPERA experiment on superluminal $m^2<0$ neutrinos being correct.\cite{OPERA}  
Given the reported OPERA excess over light speed, one can calculate the muon neutrino mass that would give that result.  Using the average neutrino energy in the experiment $E=17 GeV$ one find a  superluminal neutrino masss:

\begin{equation}
m^2=E^2(1 - v^2/c^2)=-14,300 MeV^2/c^4
\end{equation}

Since OPERA only measures a single average mass even if several mass eigenstates are present, this need not equal the mass of one single eigenstate. However if this mass did represent that of a single superluminal very heavy eigenstate, a SN 1987A  neutrino having this mass and an energy $E\approx 20 MeV$ it would have reached Earth in the past by:

\begin{equation}
t - t_0 = \frac{L_0}{c} (\frac{1}{\sqrt{1-m^2/E^2}}-1 )\approx -135,000y
\end{equation}

which could explain the absence of this eigenstate in the SN 1987 data, i.e., why the OPERA result need not be in conflict with the more stringent upper limit from SN 1987A on the value of $(v - c)/c$.

Furthermore, the OPERA result, if correct, offers a way to understand why the upper limit on the mass of the electron neutrino from tritium decay: $m< 2 eV/c^2$ need not be in conflict with the "implausibly high" values of $m_H$ and $m_L$ reported here.  For example, suppose that the electron neutrino consisted of a superposition of a tachyonic and tardyonic mass states: $|\psi>=sin\theta| \psi^+> +\cos\theta |\psi^->$ where the superscripts refer to the signs of the respective $m^2$ values for the two states whose masses are $m^2=m_+^2>0$ and $m^2=-m_-^2<0$  To find the mass of the mixed state $|\psi>$ we need the expectation value of $m^2=E^2-p^2$, i.e.,

\begin{equation}
m^2 = sin\theta<\psi^+|+cos\theta<\psi^-|(E^2-p^2)\sin\theta|\psi^+>+\cos\theta|\psi^->
\end{equation}

If we assume that the mass states have a common momentum $p$ their respective energies are $E_+^2=p^2+m_+^2$ and $E_-^2=p^2-m_-^2$ where we obviously have $E_+^2>p^2>E_-^2$.  Thus, Eq. 7 gives a mass for the mixed state:

\begin{equation}
m^2 = sin^2\theta E_+^2+cos^2\theta E_-^2-p^2
\end{equation}

It is clear that $m^2$ could be anything between the superluminal (tachyonic) $m_-^2$ and the tardyonic $m_+^2$ depending on the values of the mixing angle $\theta.$  In particular, under this scenario, the large superluminal mass eigenstate implied by OPERA could be consistent with a value for the electron neutrino mass very close to zero and superluminal depending on the mixing angle.  In fact, there is previous work that supports the idea that $m_{\nu,e}^2\approx -0.25^2 eV^2/c^4$.\cite{Ehrlich1, Ehrlich2, Ehrlich3}  

In the same manner, the possibility of a very heavy superluminal mass eigenstate need not conflict with the tight limits from cosmology and astrophysics on the sum of the three neutrino masses assuming they have $m^2>0$, \cite{Abazajian}, nor would it necessarily conflict with a suggested tachyonic mass limit of $-0.33^2 eV^2/c^4$ that applies to the flavor eigenstates.\cite{Davies}  Given the much greater sensitivity of today's neutrino detectors compared to those of 1987, were there to be a once-in-a-century occurrence of a supernova in our galaxy, it could offer either a clear confirmation or refutation of the existence of the two mass eigenstates claimed here.  Furthermore, in the event of a confirmation, it would be exceedingly difficult to see how the mass limits from tritium beta decay and cosmology could be satisfied without the existence of a superluminal mass eigenstate.

\section{Acknowledgments}

I am grateful to my colleague Bob Ellsworth for helpful suggestions.


\begin{thebibliography}{00}
\bibitem{Number} This number is often cited as 11, but we have follwed the practice of reference 5 of not dropping event 6.
\end{thebibliography}
\end{document}